\documentclass[11pt,english,a4]{scrartcl}
\usepackage{cite}
\usepackage{amsmath,amssymb,amsfonts}
\usepackage{algorithmic}
\usepackage{graphicx}
\usepackage{textcomp}
\usepackage{xcolor}
\usepackage{url}
\usepackage{comment}
\usepackage[margin=1in]{geometry}

\title{Detection of broadband squeezed light with a low-noise die-level balanced receiver 

\thanks{This research was funded in part by the Austrian Science Fund (FWF) [Grant number I5656-N] and by the Austrian Research and Promotion Agency (FFG) through project QITTY [Grant number 45004305]}
}
\date{}

\begin{document}
\maketitle

\begin{abstract}
\centering{\large{ Emmily Zaiser, Alessandro Trenti, Dinka Milovančev, Nemanja Vokić, Bernhard Schrenk, Hannes Hübel\\
  \textit{AIT Austrian Institute of Technology, Center for Digital Safety \& Security, Vienna, Austria}\\
  Corresponding author: emmily.zaiser@ait.ac.at}}\\
\end{abstract}
\par
\par
\begin{abstract}
The generation and detection of squeezed light through spontaneous parametric down-conversion in a nonlinear crystal up to a frequency of 3.5\,GHz is presented. We characterize the quantum state with balanced homodyne detection, leveraging a low-noise die-level receiver. 
\end{abstract}

\section{Introduction}
Squeezed light has been a subject of extensive research in quantum optics since the first successful experiment in 1985 \cite{first}. It  is characterized by a reduced uncertainty in one quadrature of the electromagnetic field. The generation of squeezed states is of particular interest due to applications in quantum computing, communication and high-precision measurements like the detection of gravitational waves \cite{Gravitational waves}. Photons stand out as ideal qubits due to their resistance to decoherence, crucial for large-scale quantum computing.Remarkably, quantum supremacy was claimed in 2020 in a Gaussian Boson sampling demonstration, which exploited squeezed light as a quantum resource \cite{Computing}. 
There are various approaches for the generation of squeezed light states \cite{Generation}. In this work, we achieve this through spontaneous parametric down-conversion (SPDC) in a nonlinear crystal, specifically a waveguide-based periodically poled Lithium niobate (LiNbO$_{3}$) crystal. This generates quadrature squeezed vacuum states in the process of three-wave mixing. For the characterization of the state balanced homodyne detection (BHD) is used. Instead of using bulky commercial balanced detectors with limited bandwidth and noise performance, we detect the squeezed light states with a self-made integrated die-level balanced receiver \cite{Dinka} at GHz frequencies. The receiver is composed of a low noise transimpedance amplifier (TIA) wire-bonded to two small photodiodes (PDs) in a balanced configuration which can be accessed either by a planar lightwave circuit (PLC) chip or bare fibers. The parasitic effects are minimized by using an on-chip solution for both the electronic and photonic elements and thereby achieving low-noise operations. Interfacing with a PLC chip further minimizes the footprint, moving away from fiber based splitters to a more integrated chip-based solution. 
The results of this work contribute to the ongoing efforts in developing solutions with a high level of integration for quantum computing and communication. 

\section{Experimental setup}
\begin{figure*}
    \centering \vspace*{-5mm} 
    \includegraphics[width=1\textwidth]{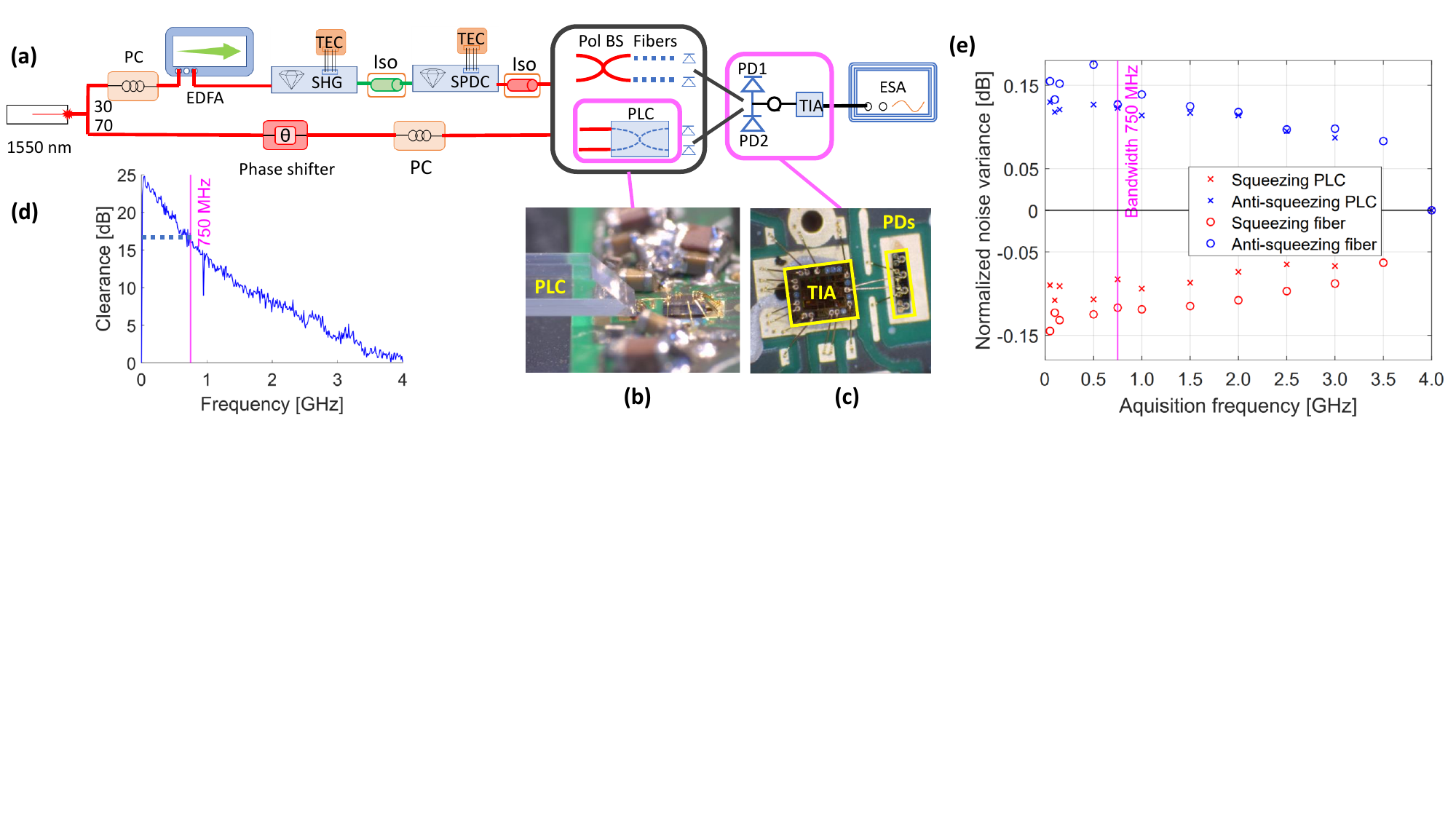} \vspace*{-30mm}
    \caption{(a) Setup for squeezed light generation and detection, (b) PLC coupled to on-chip receiver, (c) PD array chip wire-bonded to a low-noise TIA, (d) Clearance measured at $P_{\text{LO}}=$ 3.6\,dBm LO power against Frequency of the die-level receiver, (e) Estimated squeezing and anti-squeezing values at different acquisition frequencies for fiber coupled and PLC receiver setup. Each point corresponds to the average of the quadrature noise power normalized to shot noise level.}
    \label{fig: Squeezing}
\end{figure*}

For the generation and detection of squeezed light a setup as depicted in Fig. \ref{fig: Squeezing}(a) is used. A 1550\,nm laser with low linewidth of 1\,kHz is split by a 70:30 beam splitter (BS). The 30\,\% channel is coupled into a polarization controller (PC) to adjust the correct polarization for the type-0 nonlinear process happening in the crystals. The power is enhanced to 20\,dBm by an an erbium-doped fiber amplifier (EDFA) and fed into the first crystal dedicated to second harmonic generation (SHG). The SHG crystal frequency doubles the 1550\,nm light into 775\,nm light which is used to pump the SPDC crystal and generate broadband squeezed light with a center wavelength of 1550 nm and full width at half maximum of 152 nm. The efficiency of both, SHG and SPDC, are temperature dependent due to the phase matching condition. Two temperature controllers are used to optimize the phase matching and maximize the output powers of the SHG and SPDC crystals. To avoid leakage of the pump beam two isolators are included after the nonlinear crystals. The power of the visible beam coupled into the SPDC crystal is at $P_{\text{in}}=-12.0$\,dBm. The 70\,\% channel of the 1550\,nm laser is used as local oscillator (LO) and modulated by a phase shifter at 200\,Hz to sweep the quadrature variances. A PC is used to optimally align the LO polarization, resulting in $P_{\text{LO}}=$ 3.6\,dBm before mixing with the squeezed light. 
We investigate two different approaches and compare the results. One in which 
two separate bare fibers are aligned over the PDs of the die-level BHD receiver, guiding the mixed beam. For this, a 3\,dB polarization dependent beam splitter (Pol BS) is implemented. A more compact solution is offered by utilizing a PLC that functions as 3\,dB splitter and facilitates vertical coupling through its 42$^{\circ}$-angled output facet as can be seen in Fig. \ref{fig: Squeezing}(b). The electrical output of the receiver is measured with an electrical spectrum analyzer (ESA).
This allows to directly measure the field quadrature variance \cite{Loudon}
\begin{equation}
    (\Delta X_{\beta})^{2} =\eta (e^{2r}\cos^{2}{\Theta}+e^{-2r}\sin^{2}{\Theta}) +1-\eta
    \label{eq: Quadratures}
\end{equation}
wherein $\Theta$ is the LO phase, $\eta$ the overall detection efficiency and $r=\mu\cdot\sqrt{P_{\text{in}}}$ the squeezing parameter wherein $\mu$ is a parameter proportional to the crystal length and the nonlinear interaction strength. By changing the phase of the local oscillator, we can directly access the field quadrature variances. For $\Theta=\pi/2$ the amount of squeezing can be estimated while the LO phase of $\Theta=\pi$ gives the anti-squeezing. The receiver consists of an array of top-illuminated die-level PDs which are wire-bonded in a differential configuration to an ultra low-noise TIA as shown in Fig. \ref{fig: Squeezing}(c). The PDs have a high responsivity of 1.1\,A/W, presenting a satisfying efficiency of 88\,\%. The detector has a high clearance $cl$, especially considering the low LO power $P_{\text{LO}}=$ 3.6\,dBm, within the bandwidth of 750\,MHz that lies between $cl$ = 25\,dB for low frequencies and $cl$ = 16\,dB at 750\,MHz. The clearance $cl$ is still high at frequencies near 750\,MHz and decreases steadily until it reaches a plateau at 4\,GHz with $cl$ = 0.2\,db as can be seen in Fig. \ref{fig: Squeezing}(d). The common mode rejection ratio is 35\,dB and the noise equivalent power $<2$\,pW/$\sqrt{\text{Hz}}$. Those characteristics promise a good performance in the detection of squeezed light states even at frequencies above the receiver's 3\,dB bandwidth.

\section{Squeezed light generation and detection}
The squeezing is broadband, meaning we can choose where to measure within a vast frequency range. It is interesting to investigate how the amount of (anti-) squeezing changes with the frequency. We expect lower squeezing values at higher frequencies. This is related to the electrical noise that increases at higher frequencies because of a lower clearance. According to the parameter $\eta_{\text{el}}=(cl-1)/cl$, this leads to a lower efficiency. To testify  this, the amount of (anti-) squeezing ($M_{\text{A-Sq}}$) $M_{\text{Sq}}$ is evaluated by averaging over the (maxima) minima of the noise power normalized to the shot noise level.
We calculate this at different acquisition frequencies, using an ESA, acquiring over 50\,ms with a resolution bandwidth of 20\,MHz and a video bandwidth of 1\,kHz. The estimated (anti-) squeezing values against the center frequency are plotted in Fig. \ref{fig: Squeezing}(e) for both the fiber coupled and PLC setup. 
As expected, the measured amount of squeezing decreases with the increasing frequency. It is not possible to detect squeezing above 3.5\,GHz due to the low clearance. From the experiment conducted with the fiber coupling, we measure the highest amount of squeezing $M_{\text{Sq}}=$ -0.15\,dB and anti-squeezing $M_{\text{A-Sq}}=$ 0.16\,dB at the center frequency of 50\,MHz. This corresponds to -3.07\,dB of squeezing and 3.08\,dB of anti-squeezing at the SPDC crystal, taking into account the connection losses, the detection efficiency, the electrical noise, and the output coupling of the crystal, yielding $\eta=0.51$. The value is comparable to other state-of-the-art experimental results \cite{Optica}. While offering the advantage of a lower footprint and higher level of integration, using the PLC gave lower squeezing amounts. This is due to slightly reduced coupling efficiency of the PLC and possibly polarization issues since the PLC is not polarization maintaining.
\newline In conclusion, we have implemented a fully wave-guided setup for squeezed light generation and detection, utilizing a die-level receiver at frequencies up to 3.5\,GHz. This represents an important step in developing scalable solution for quantum communication applications. Future work will involve an improvement of the setup to achieve higher LO powers as well as an investigation of higher bandwidth receivers with improved coupling.
\section*{Acknowledgment}
We thank Florian Honz and Daniel Pereira for their research contributions by supporting us in the lab and for the various fruitful discussions that led to this paper.

\vspace{12pt}

\end{document}